\documentclass[sigconf, authorversion=true, nonacm=true]{acmart}
\AtBeginDocument{%
  }

\setcopyright{acmcopyright}
\copyrightyear{2018}
\acmYear{2018}
\acmDOI{XXXXXXX.XXXXXXX}

\copyrightyear{2023}
\acmYear{2023}
\setcopyright{rightsretained}
\acmConference[CHI '23]{Proceedings of the 2023 CHI Conference on Human Factors in Computing Systems}{April 23--28, 2023}{Hamburg, Germany}
\acmBooktitle{Proceedings of the 2023 CHI Conference on Human Factors in Computing Systems (CHI '23), April 23--28, 2023, Hamburg, Germany}\acmDOI{10.1145/3544548.3581226}
\acmISBN{978-1-4503-9421-5/23/04}

\usepackage{xac}

\begin{document}

\title{GANravel: User-Driven Direction Disentanglement in Generative Adversarial Networks}

\author{\textbf{Noyan Evirgen}}
\affiliation{%
  \institution{UCLA HCI Research}
  \country{}}
\email{nevirgen@ucla.edu}

\author{\textbf{Xiang `Anthony' Chen}}
\affiliation{%
  \institution{UCLA HCI Research}
  \country{}}
\email{xac@ucla.edu}

\begin{abstract}
Generative adversarial networks (GANs) have many application areas including image editing, domain translation, missing data imputation, and support for creative work.
However, GANs are considered `black boxes'. Specifically, the end-users have little control over how to improve editing directions through disentanglement.
Prior work focused on new GAN architectures to disentangle editing directions.
Alternatively, we propose \gr---a {\it user-driven} direction disentanglement tool that complements the existing GAN architectures and allows users to improve editing directions iteratively.
In \hl{two} user studies with \hl{16} participants each, \gr users were able to disentangle directions and outperformed the state-of-the-art direction discovery baselines in disentanglement performance. \hl{In the second user study, \mbox{\gr} was used in a creative task of creating dog memes and was able to create high-quality edited images and GIFs.}\blfootnote{To appear in CHI 2023}
\end{abstract}



\begin{teaserfigure}
\centering
  \includegraphics[width=\textwidth]{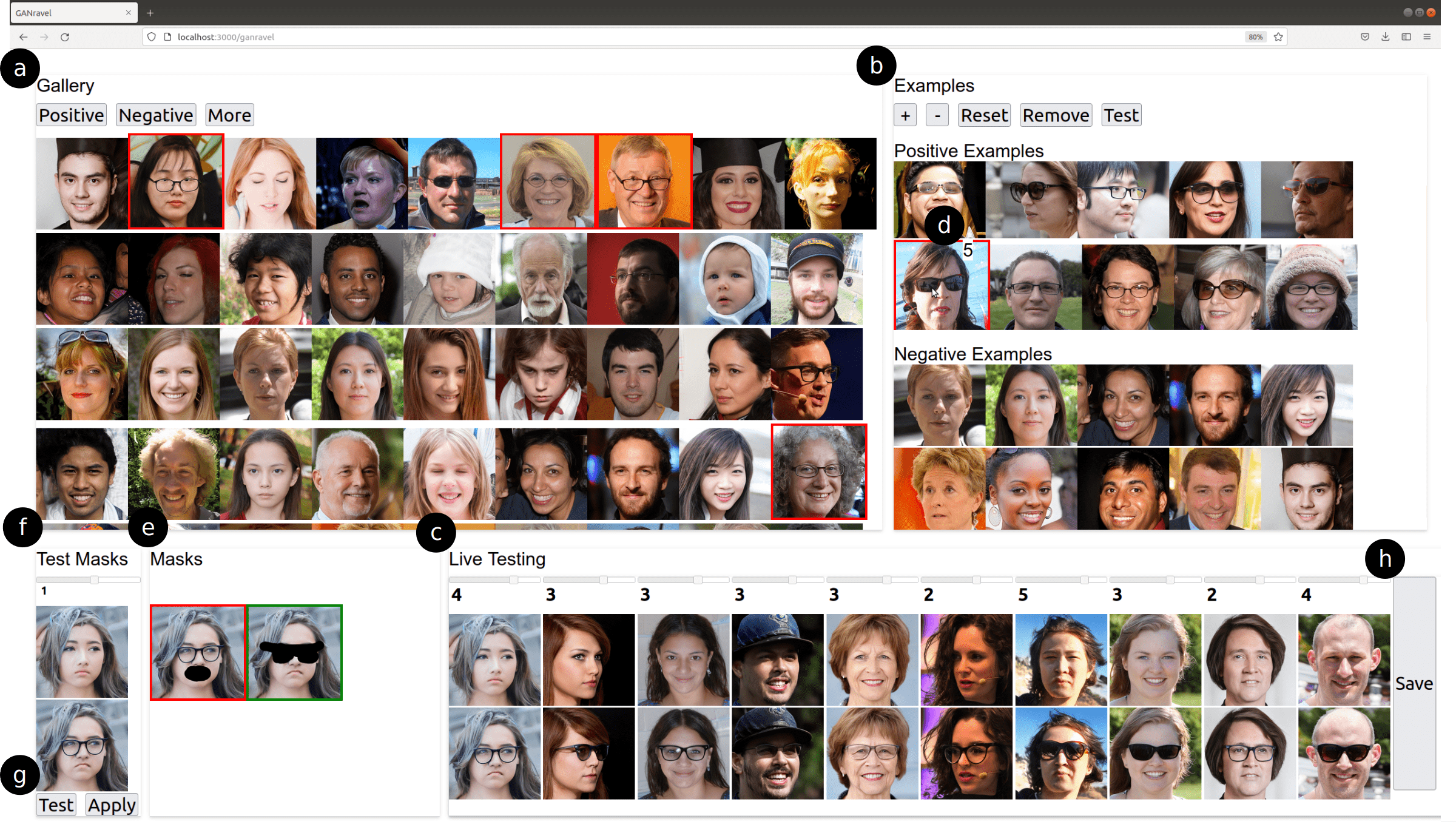}
  \caption{\gr enables users to disentangle editing directions in generative adversarial networks (GAN) using global and local disentanglement approaches. 
  (a) A direction is often entangled when created by selecting exemplary images from the gallery.
  (b) The weights of the exemplary images can be adjusted to disentangle global attributes such as age and gender. (c) The direction can be tested on the live-testing section using multiple test images. (d) The user can hover over an exemplary image to see its weight and go back and forth between weight adjustments and live-testing until global attributes are disentangled. (e) The user can use masks to disentangle local attributes such as glasses and closed mouth. (f) The masks can be combined to either preserve or discard a region of interest and they can be tested. (g) Resulting disentangled direction can be applied to other test images in the live-testing section. (h) The final disentangled direction can be saved and applied in other future images.}
  \label{fg:fig1}
\end{teaserfigure}

\maketitle

\section{Introduction}


Generating complex data is a long-standing problem in computer sciences. 
In recent years, generative adversarial networks (GANs) are shown to be suitable for such tasks including medical imaging~\cite{yi2019generative}, data enhancement~\cite{pascual2017segan, wang2018esrgan} and image editing~\cite{zhu2016generative}. Moreover, human-AI collaboration is shown to be useful in various areas including medicine~\cite{10.1145/3577011, 10.1145/3449084} and art creation~\cite{mateja2021towards}.
Unfortunately, GANs function as `black boxes' by their nature. As a result, end-users have little control over the generative process which limits human-AI collaboration. 
When an end-user with creative intent uses a GAN model for editing, the lack of control over the capabilities of the model can lead to inadvertent and inconsistent results.
Moreover, there is little support for the end-users to \textit{improve} said controls (\textit{editing directions}~\footnote{simply referred to as `direction' in the paper})-- a set of input parameters that steers the GAN model toward the intended characteristics with varying levels (e.g. making the eyes on the input image bigger or smaller while preserving other characteristics).
Specifically, a common problem is that directions can be \textit{entangled}, \ie while the direction changes the desired attribute, it might change other unintended attributes as well (\eg the direction that adds glasses to the people can also change their gender). Entanglement can also result in a direction that only works on a certain type of image (\eg the direction that adds glasses to the people, may not be able to apply on young people).

Improving an entangled direction, \ie \textit{disentanglement}, is an active research area in GAN research.
InterFaceGAN~\cite{shen2020interfacegan} finds entangled directions using pre-trained classifiers and disentangles them via subspace projection. However, InterFaceGAN requires developers to come up with disentanglement rules for each domain and use human annotators to label large datasets. There are also GAN architectures that have improved disentanglement properties over the traditional GAN \cite{goodfellow2020generative} such as StyleGAN~\cite{karras2019style}, and GANformer~\cite{hudson2021generative}. StyleGAN achieves better `attribute separation' by introducing `styles' which is inspired by the style transfer literature. GANformer has better `spatial decomposition' since it leverages transformers~\cite{vaswani2017attention}.
Despite their promises, these algorithm-driven solutions cannot capture a user's intent of disentangling specific attributes of a given image; thus it is important and complementary to support a {\it user-driven} approach, especially when algorithms fail to achieve their desired disentanglement effects. 

To this end, we design and implement \gr, a tool that allows users to interactively and iteratively disentangle directions. 
\hl{In order to highlight how \mbox{\gr} can enable users to disentangle directions, we focus on two studies: \one the common task of editing human faces and \two the creative task of creating memes of dogs. \mbox{\gr} can use any image-generating GAN model that has disentangled directions. We complement different GAN models --StyleGAN2~\cite{karras2020analyzing} and FastGAN~\cite{liu2020towards}-- in the studies to underline the model-agnostic nature of \mbox{\gr}.}

\gr achieves disentanglement through two main approaches, \textit{global} and \textit{local} disentanglement. Global disentanglement focuses on the holistic attributes (\eg gender, age, lighting, etc.) and aims to disentangle a direction by balancing these attributes. This is achieved by users' selecting \textit{exemplary} images that carry the entangled attribute and adjusting the weights of these images. Complementarily, local disentanglement focuses on attributes of specific components (\eg hair style, eye size, smiling, etc.) which can be more subtle than the global attributes. Therefore, they are inherently harder to balance by tuning exemplary images. Instead, users disentangle local attributes through masking where the user highlights a region of entanglement. Then, the masks are used to find the entangled attribute in a one-shot manner (from a single mask without training). 

To validate \gr, we conducted \hl{two user studies with 16 participants each. The first user study had coarse- and fine-grained tasks. The second user study had three tasks where we compared \mbox{\gr} with state-of-the-art user-driven direction discovery method \mbox{\gz}. In the first user study,} coarse-grained tasks were adding glasses, making the face smile more, and making the face older. Fine-grained tasks were adding lipstick, making eyes bigger, and making hairs curlier. 
Participants were asked to find the target direction. Results show that participants were successful in finding disentangled directions. Specifically, in the coarse-grained tasks, participants found directions that were more disentangled than the directions that are found with state-of-the-art methods. Disentanglement was measured with two main analyses: \one facial identity preservation and \two facial attribute classifier based metric. 
Results show that our metrics were `similar' across tasks and participants felt satisfied with their edits and disentanglement performance. Participants' perceived disentanglement success was also aligned with our iterative disentanglement metrics which showed that a directions got more disentangled as participants iterated it more with \gr. \hl{In the second user study, participants were tasked to create dog memes by finding disentangled directions. Participants used \mbox{\gr} to disentangle directions they found using \mbox{\gz}. Disentanglement was measured through a classifier based metric. Results show that participants were able to disentangle directions which was also aligned with their perceived success.}  


\gr makes a \textbf{tool contribution}, enabling a user to interactively and iteratively disentangle a direction and improve generated results. \gr provides two user-driven approaches for disentanglement, encompassing global and local disentanglement. Overall, \gr complements existing GAN architectures and the user study showed that resulting directions were more disentangled compared to the state-of-the-art direction discovery methods.

\section{Background \& Related Work}

In this section, we first review background information about GAN, editing directions, and entanglement. Then, we review two areas of prior work: \one existing algorithm-driven approaches for direction discovery and \two existing approaches that enable users to interact with GAN.

{\bf Generative adversarial networks}
As shown in \fgref{gan}, a typical GAN \cite{goodfellow2020generative} consists of two neural network models: the generator ($G$) and the discriminator ($D$). $G$ is trained to generate synthetic data $x_g$ from a data domain such as human faces. At each training step, the goal of $G$ is to `trick' $D$ by creating $x_g$ that is indistinguishable from the data from the aforementioned data domain ($x_r$), and the goal of $D$ is to distinguish synthetic data $x_g$ from the real data $x_r$. By playing this zero-sum game, both networks are trained based on the prediction of $D$ until $G$ is trained to generate realistic data. By default, $G$ generates $x_g$ from a randomly-sampled noise vector ($z$). GAN functions as a `black box' because the space where $z$ resides is considered to be highly nonlinear. As a result, end users have very little control over the generative process.

\fg{gan}{gan}{1}{A typical GAN model and its' training scheme. Dashed lines indicate the gradients that train the generator ($G$) and discriminator ($D$). In this training step, $G$ failed to trick $D$ and the weights of $G$ are updated accordingly. $z$ controls the generated data $x_g$. The direction ($d$) is defined as the vector that changes the input $z$ with addition. When $d$ is normalized, $\lambda$ is referred to as the strength of the direction.
}

{\bf GAN editing direction}
Formally, the generator learns a mapping function $f: Z \mapsto X$ where $Z \in \mathbb{R} ^ {n}$ and $X$ is the space of the data domain. $n$ is the dimension of the input vector which depends on the generator model. $Z$ is referred to as the `latent space'. Typically, when $G$ is trained, latent space is sampled from a Gaussian distribution \cite{goodfellow2020generative}. The resulting vector ($z$ in \fgref{gan}) can be moved in the latent space to change the output in a semantically meaningful way along the editing direction ($d$ in \fgref{gan}). For example, given a face without glasses $f(z_0)$, the initial vector can be edited so that the output $f(z_1)$ is the same face as $f(z_0)$ but wearing glasses, where $z_1 = z_0+\lambda d$. The coefficient $\lambda$ is the `strength' of the edit when $d$ is normalized.

{\bf Entanglement in editing directions}
Entanglement is when a direction $d$ changes multiple semantic attributes of the output simultaneously. For example, if $f(z_0)$ outputs a face without glasses, and $f(z_0+\lambda d)$ outputs the same face but with glasses and curlier hair, then the facial attributes `glasses' and `curly hair' are considered to be `entangled' for the direction $d$. In this example, while trying to edit the face to add glasses, another attribute is unintentionally changed. Entanglement can happen through the direction discovery process or the generative model can introduce entanglement due to bias. \textit{Disentanglement} is the process of improving the entangled direction, resulting in a direction that only changes the intended attributes.

{\bf Algorithm-driven discovery in GAN}
There are many supervised direction discovery approaches that require a large labeled dataset \cite{jahanian2019steerability, Goetschalckx_2019_ICCV, Shen_2020_CVPR, plumerault2020controlling, yang2021semantic, radford2015unsupervised, tzelepis2021warpedganspace}. For example, InterFaceGAN finds disentangled directions by using classifiers and subspace projection~\cite{Shen_2020_CVPR}. The downside of these approaches is that they rely on attribute predictors or human annotators. \hl{Relying on predictors or annotators severely limits the number of directions, while \mbox{\gr} provides arbitrary range of directions through disentanglement.} Recently, there has been some interest in finding directions in an unsupervised manner. Voynov \etal finds directions in an unsupervised manner by training a `reconstructor' that predicts the strength and the index of a randomly sampled direction~\cite{pmlr-v119-voynov20a}. SeFa reveals underlying variation factors in an unsupervised manner using closed-form factorization~\cite{Shen_2021_CVPR}. GANSpace finds unsupervised directions using PCA on the feature space~\cite{harkonen2020ganspace}.
Collins \etal applies k-means to the hidden layer activations of the generator to find a decomposition of the generated output into semantic objects. Then, the generative model is able to transfer a style of a facial attribute in one image to another image using the decomposition and the respective style parameters~\cite{Collins_2020_CVPR}. Although there are many algorithm-driven direction discovery methods, there is little support for improving the directions through disentanglement. \gr enables users to disentangle a given direction with simple interactions.

{\bf Enabling users to interact with GAN}
Interactions with generative adversarial networks have been an active research area. Typically, the users interact with GANs through sliders similar to \gr. \hl{We decided to use sliders in the live-testing area following prior work~\cite{shen2020interfacegan, abdal2021styleflow}.} Dang \etal compared the regular sliders and sliders that provide feedforward information (`filmstrips') in a comparative study \cite{dang2022ganslider}. Zhang \etal enabled users to explore the latent space with a grid-like view of sampled images \cite{10.1145/3411764.3445714}. The users can zoom in or out, pivot, and pan to explore the latent space. \hl{Even though the gallery section of \mbox{\gr} is not spatially meaningful, it also leverages a grid of images for selecting exemplary images}. Chiu \etal created a tool that allows the user to search through the GAN latent space interactively with a one-dimensional slider~\cite{chiu2020human}. There are also various prior works that can enable new user interactions. Heim showed that GANs can iteratively accept inputs to `generate an image more like A than B'~\cite{heim2019constrained}. Chen \etal showed that it is possible to generate human faces from human face sketches~\cite{chen2020deepfacedrawing}. Cheng \etal developed a visual design assistant that interacted with the users through natural language and edited the GAN outputs~\cite{10.1145/3394171.3413551}. Ling \etal allows users to edit images leveraging segmentation masks of the images~\cite{ling2021editgan}. StyleCLIP accepts a textual description of the direction and finds it using the CLIP model~\cite{Patashnik_2021_ICCV, radford2021learning}. \gz allows users to discover directions via iterative scatter/gather interactions and complements other direction discovery methods~\cite{evirgen2022ganzilla}. \gz is a `complementary' solution to the algorithm-driven discovery methods. It does not find a target direction or improve disentanglement, whereas \gr is a user-driven disentanglement tool that \hl{provides more flexibility to disentangle user-defined directions}. \hl{\mbox{\gz} uses thumbnail images to represent directions and uses a brush tool to highlight a region of interest which is similar to how users navigate in \mbox{\gr} and highlight a mask for local disentanglement}.

\section{Design \& Implementation}

In this section, a detailed walkthrough of \gr's design and implementation is given using an exemplary use case of adding glasses to the face. \hl{Similar to GAN Dissection~\cite{bau2018gan}, \mbox{\gr} uses the \textit{filters}\footnote{referred to as neurons or units in~\cite{bau2018gan} interchangeably.} in GAN models to edit images. Specifically, we extract feature maps from each filter and define the directions at the filter level. Specific to StyleGAN, we define the directions in StyleSpace.} StyleSpace refers to the space of style parameters that scale the outputs of convolutional filters of StyleGAN2. StyleSpace is considered to be more disentangled than the latent space which makes it suitable for \gr~\cite{Wu_2021_CVPR}. 
\hl{For the other GAN models, we define a space similar to StyleSpace. Each filter has a value associated to it which is found by averaging their feature map. When all of the associated values are concatenated, it creates a vector which we define as the direction. In other words, each dimension of the direction represents a filter in the GAN model similar to StyleSpace. These directions can then be applied to other reference images similar to \cite{bau2018gan}, where the feature maps are increased or decreased by the respective dimension of the direction.}
\gr consists of two main disentanglement approaches, global and local disentanglement which are detailed in \S~\ref{subsec:global} and \S~\ref{subsec:local} respectively.

\fg{select}{select}{1}{Exemplary image selection in \gr. (a) The positive and negative examples can be selected from the gallery. (b) Users can request more images. (c) The resulting direction can be tested on test images using the `test' button. (d) The weights of the examples can be changed using the `+' and `-' buttons.}

{\bf Starting with an entangled direction} The workflow of \gr starts with an entangled direction which can be achieved by selecting a handful of positive and negative exemplary images from the image `gallery' (\fgref{select}a). 
Users can also request more images (\fgref{select}b). 
Positive examples carry the target attribute (glasses) whereas negative examples do not. After the selection is complete, the current entangled direction can be tested in the `live-testing' area (\fgref{fig1}c) using the `test' button (\fgref{select}c).

\subsection{Global Disentanglement}
\label{subsec:global}

The edited test image with the entangled direction can be seen in \fgref{global_local_disentanglement}a. As it can be seen, the glasses direction is entangled with age (person gets older). 
In order to disentangle glasses direction from age, the weights of exemplary images need to be adjusted which can be done by the `+' and `-' buttons (\fgref{select}d). The weights can be seen when the mouse hovers over the images (\fgref{select}e). Since the entangled attribute is age, the weights of young positive images are increased (\fgref{global_local_disentanglement}b). Another approach is to increase the weight of old negative images. \hl{In an earlier pilot study, we discovered that} global disentanglement requires back and forth between weight adjustments and live-testing. It is also beneficial to look for new exemplary images (\eg positive images that are young and have glasses). After changing the weights to balance out the entangled feature age, the edited test image with the globally-disentangled direction can be seen in \fgref{global_local_disentanglement}d. As it can be seen, the resulting glasses direction is disentangled from age, but still has some entanglement issues (\eg mouth is closed, beard, etc.). These subtle issues are harder to disentangle with positive/negative examples and weight adjustments. They are instead fixed with local disentanglement.  


\fg{global_local_disentanglement}{global_local_disentanglement}{1}{\gr have two disentanglement approaches, global and local disentanglement. (a) The resulting image when an entangled direction (glasses entangled with age and mouth) is applied to the reference image. (b) Global disentanglement. The weights of the young images with glasses are increased to disentangle age. (c) Local disentanglement. The area of the intended direction (glasses) is preserved while the area of the entangled attribute (mouth) is discarded. (d) The resulting image when the globally-disentangled direction is applied to the reference image. Age is disentangled but mouth is still entangled. (e) The resulting image when the locally-disentangled direction is applied to the reference image. Mouth is disentangled but age is still entangled.}

{\bf Implementation} First, \hl{the vector (associated with the filters)} of each selected image are extracted. The weighted average vector of positive examples is subtracted from the weighted average vector of negative examples, resulting in the discovered direction. Initially, all the positive examples have a weight of 1 and all the negative examples have a weight of -1. By increasing the absolute value of positive or negative weights, the user can increase the effect of individual vectors or vice versa. Since the user only selects a handful of images, it is not required for the user to select negative examples to save time and effort. Instead, each selected vector is subtracted by an `average' vector which is created by averaging 10000 vectors that are extracted from random image samples. As a result, even if the user only selects positive examples, the difference between selected vectors and average vector carries enough information to find the target direction.

\subsection{Local Disentanglement}
\label{subsec:local}

In addition to global disentanglement, \gr can also disentangle local attributes. This is achieved by double clicking the test image with entangled attributes which pops up a brushing tool for highlighting. 
After the region of interest is highlighted, the mask will appear in the `masks' section (\fgref{fig1}e). 
The masks can be used in two ways: \one to `discard' the attribute and \two to `preserve' the attribute.
The discard feature changes the direction to be \textit{less} affected by the masked area and the preserve feature allows the direction to be \textit{more} affected by the masked area. 
For example, considering the same entangled direction in \fgref{global_local_disentanglement}a, it can be seen that glasses direction is entangled with not only age (global) but also closed mouth (local).
The masks that are used to `preserve' the glasses and `discard' the closed mouth can be seen in \fgref{global_local_disentanglement}c.
The edited test image with the locally-disentangled direction can be seen in \fgref{global_local_disentanglement}e.
As can be seen, the edited image still has the glasses which is achieved by the `preserve' mask.
Moreover, the mouth is no longer closed similar to the reference image which is achieved by the `discard' mask. 
However, the direction is still entangled with age which is a global entanglement issue.
The masks can be tried in combination by selecting them and clicking the `test` button (\fgref{fig1}g). Each click cycles through green, red, and no selection for the masks. If the user wants to test the masked direction on all the test images, the `apply' button can be used (\fgref{fig1}g). Finally, the directions can be saved with the `save' button (\fgref{fig1}h).

{\bf Implementation}. When a region is highlighted, the filters that are responsible for that region can be found. We extract all the outputs of the filters as \hl{feature maps}. We then calculate the overlap between the images and the mask, scaling the mask for each layer. After normalizing the overlap, we use the overlap value as the `importance' metric for that particular filter. Higher values indicate that the \hl{filter} is highly `responsible' for the region of interest. For the `preserve' feature, we scale \hl{the vector} with the importance metric which results in more `important` filters having stronger influences on the output. For the `discard' feature we use the same importance metric but we inversely scale \hl{the vector} which results in more `important' filters having weaker influences on the output. This simple yet effective trick allows us to disentangle a direction in a one-shot manner without any training or data in real-time.

\subsection{Other Implementation Details}
\label{subsec:other}
We used Nvidia's implementation of StyleGAN2 in PyTorch and their pre-trained model \hl{for the first user study. We trained a FastGAN model for the second user study. We extended the PyTorch code of StyleGAN2 and FastGAN to be able to extract style parameters (only in StyleGAN2) and filters which are used in disentanglement.} We used Python for all of the back-end calculations and Flask for our web-framework. The front-end of \gr was developed on Javascript, Node.js, and React. The back-end ran on a Linux 18.04 server which is equipped with an Nvidia GeForce RTX 3090 GPU. 

\section{User Studies}
\hl{We conducted two user studies to validate whether \mbox{\gr} can enable users to disentangle directions that edit images for \textit{creative purposes}. 
In the first user study, we asked participants to find and disentangle directions for human faces using \mbox{\gr} in two sets of tasks: \one coarse-grained and \two fine-grained tasks. 
The goal of the first user study is to analyze the \textit{disentanglement performance} of \mbox{\gr} with respect to the state-of-the-art methods on the standard face editing task. 
In the second user study, we asked participants to create dog memes using \mbox{\gr}, \mbox{\gz} and combination of both.
The goal of the second user study is to put emphasis on the tool contribution of \mbox{\gr} by enabling users in a creative task of generating memes. The second user study also shows how \mbox{\gr} can disentangle directions that are found by another direction discovery method such as \mbox{\gz} which is the most similar work to ours.
}

\subsection{Editing Human Faces}
\hl{The details of the first user study is provided below.}

\textbf{Participants.} We used convenience sampling to recruit \hl{16} participants from a local university. \hl{Out of 16 participants, twelve were male, four were female, and they were aged from 23 to 33}. 
\hl{Seven participants majored in electrical engineering, two in biomedical engineering, five in economics, and two in computer science.}
All of the participants had programming experiences from five to 10 years and none of them had programmed or used GAN-enabled applications before.

\textbf{Tasks \& Procedure.} 
Each participant performed two sets of tasks (coarse-grained and fine-grained tasks) using \gr, and each task consists of three trials. In both tasks, the participant's goal was to use \gr to find a direction that steers the output of the GAN towards the given editing goal while preserving other attributes.

\begin{itemize}
    \item \textbf{Coarse-grained tasks.} In each trial, participants were given a generic editing goal. Specifically, the goals were: adding glasses to the faces, making the faces smile more and making the faces appear older\footnote{Hereafter simply referred to as glasses, smile and age}. The goals were intentionally generic so that they can be compared with the state-of-the-art face editing methods.
    \item \textbf{Fine-grained tasks.} In each trial, participants were given a specific editing goal. Specifically, the goals were: adding lipstick to the faces, making eyes bigger, and increasing the curliness of the hair\footnote{Hereafter simply referred to as lipstick, eye, and curliness}. The goals were intentionally specific which are not naturally supported by the prior works to show the flexibility of \gr. 
\end{itemize}
Each user study started with an introductory tutorial of \gr. After the tutorial, the participants are given a brief practice session to try out \gr using a toy example. We then continued with a block of tasks (either coarse- or fine-grained) which is followed by a short break. After the break, the participants are given the remaining block of tasks. The order of the tasks and the three trials within each block were counter-balanced across participants. We concluded the study with a semi-structured interview to elicit participant's qualitative feed-back of \gr. The entire study took place over Zoom and lasted for about an hour. 
Each participant was compensated with a \$25 gift card.

\textbf{Data \& Apparatus.}
We used StyleGAN2 as our GAN model, specifically we used a pre-trained model which is trained on Flickr-Faces-HQ (FFHQ) dataset. The model parameters and its PyTorch code are available on Nvidia's github page\footnote{\href{https://github.com/NVlabs/stylegan2-ada}{https://github.com/NVlabs/stylegan2-ada}}. Other implementation details are given in \S~\ref{subsec:other}. The user studies are conducted virtually over Zoom. Each participant used Zoom's remote control feature to interact with the computer of the experimenter. \gr ran on the same computer to minimize latency. 

\subsection{Generating Dog Memes}
\label{section:cat-memes}
\hl{The details of the second user study is provided below.}

\hl{
\textbf{Participants.} 
We used convenience sampling to recruit 16 participants from a local university. The participants of the first and second user studies did not overlap. Out of 16 participants, ten were male, six were female, and they were aged from 21 to 29. Five participants majored in electrical engineering, one in industrial engineering, three in biomedical engineering, two in economics, and five in computer science.
All of the participants had programming experiences from five to 10 years and none of them had programmed or used GAN-enabled applications before.
}

\hl{
\textbf{Tasks \& Procedure.} Each participant performed three tasks using \one \mbox{\gr}, \two \mbox{\gz} and \three combination of both. Specifically, the participants were asked to generate two dog memes for each task. A meme consisted of two images side-by-side and text underneath them. The images were either the unedited reference images, or an edited image, which is generated by applying the discovered direction to the reference image. The text underneath them was written by the participants. We also let participants turn these two images into a single GIF to animate the meme~\footnote{We only reported the images in the paper. The GIFs as well as the source code will be made available on GitHub if the paper gets accepted.}.
For the first two tasks, participants generated dog memes using \mbox{\gr} and \mbox{\gz} separately. For the last task, participants disentangled the directions they found before (with \mbox{\gz}) using \mbox{\gr}. In other words, they improved the quality of the memes that they already found before. The goal in the first two tasks is to compare the disentanglement quality of \mbox{\gr} with a similar work to ours \mbox{\gz}. The goal in the last task is to show how \mbox{\gr} can be used with another direction discovery method to disentangle (improve) directions.
}

\hl{
Each user study started with an introductory tutorial of \mbox{\gr} and \mbox{\gz}. After the tutorial, the participants are given a brief practice session to try out \mbox{\gr} and \mbox{\gz} using a toy example. We then continued with a block of tasks (creating two memes using either \mbox{\gr} or \mbox{\gz}) which is followed by a short break. After the break, the participants created two more memes using the remaining tool. After the second task, participants are given another short break. After the second break, for the third task, the participants disentangled the memes (previously found with \mbox{\gz}), using \mbox{\gr}. The order of the first two tasks were counter-balanced across participants. Similar to the first user study, we had a semi-structured interview at the end. The entire study took place over Zoom and lasted for about an hour. 
}

\hl{
\textbf{Data \& Apparatus.} 
We used FastGAN~\footnote{\href{https://github.com/odegeasslbc/FastGAN-pytorch}{https://github.com/odegeasslbc/FastGAN-pytorch}} as our GAN model, specifically we used Projected GAN~\footnote{\href{https://github.com/autonomousvision/projected_gan}{https://github.com/autonomousvision/projected\_gan}} for training the model on Animal Faces-HQ (AFHQ) Dog~\cite{choi2020starganv2} dataset. \mbox{\gz} is publicly available on Github~\footnote{\href{https://github.com/noyanevirgen/GANzilla-UIST22}{https://github.com/noyanevirgen/GANzilla-UIST22}}. We adapted \mbox{\gz} for the trained FastGAN model. Other implementation details are given in \S~\ref{subsec:other}. Similar to the editing human faces user study, the user studies are conducted virtually over Zoom.
}

\subsection{Measurement}
\label{subsec:measurement}
In both user studies as the participants interacted with \gr and \gz, we saved every image generated by them. We also saved all of the user interactions, such as which buttons are clicked, and which images are selected with timestamps. We also recorded the entire session over Zoom. 

In the exit interview, first we asked participants to assess \gr based on their overall experience. Specifically, they are asked to rate (on a 7-point Likert scale) \one whether \gr is easy to use, \two whether \gr can find directions that match their editing goal, and \three whether \gr can disentangle an initially entangled direction. Next, participants rated the cognitive load using the mental demand, effort and frustration dimensions of the NASA TLX questionnaire~\cite{hart1986nasa}. Finally we asked participants to evaluate the usefulness of \gr's individual UI components: selecting positive and negative examples, changing weights of positive or negative examples, live-testing directions on multiple images, highlighting and masking to create new directions.

\section{Quantitative Results}

In this section, we provide quantitative analyses to understand participant performance and behavior using \gr and compare it with state-of-the-art baselines. There are multiple comprehensive analyses for the tasks including: 
\one disentanglement performance comparison between the user-edited images and the images edited by the state-of-the-art baselines (\S~\ref{section:disentanglement-performance}), \two further analyses into disentanglement to surface trends in entanglement (\S~\ref{section:disentanglement-patterns}), and \three user behavior (\S~\ref{section:user-behavior}). \hl{
The study of human faces in AI literature allows us to access labeled datasets, classifiers, and facial feature extraction tools. This makes it easier to analyze disentanglement performance using the first user study. We also performed analyses on the second user study, in which participants were asked to find editing directions for dogs, although these analyses were not as detailed.}

\subsection{Disentanglement performance}
\label{section:disentanglement-performance}

Measuring disentanglement accurately is an open question in GAN research. \hl{For the first user study,} we measured disentanglement through two main analyses: \one similar to \cite{khodadadeh2022latent}, we measured how well a direction preserves the facial identity, and 
\two we measured how much the intended facial attribute changed compared to unintended features using facial attribute classifiers. 
A disentangled direction should preserve the facial identity more since there are fewer facial attributes changing compared to an entangled direction. With a disentangled direction, the intended facial attribute should change more than the unintended features which can be measured with classifiers. \hl{For the second user study, we measured how well a direction preserves the breed of the dog. A disentangled direction is expected to preserve the breed more, similar to facial identity analysis in the first user study.}

\hl{\textbf{Baseline Methods}.} We quantitatively compared \gr with four state of the art methods \hl{for editing human face user study}: Inter\-Face\-GAN \cite{shen2020interfacegan}, GAN\-Space \cite{harkonen2020ganspace}, Style\-Flow \cite{abdal2021styleflow} and \gz~\cite{evirgen2022ganzilla}. 
GAN\-Space and StyleFlow are unsupervised direction discovery methods and InterFaceGAN is a supervised direction discovery method that leverages classifiers. \gz is a complementary user-driven direction discovery method. \gr does not require a dataset or a classifier similar to GANSpace, StyleFlow, and \gz. On the other hand, \gr can disentangle a given direction, similar to the conditional manipulation of InterFaceGAN. Although InterFaceGAN, GANSpace and StyleFlow do not have user interaction as their contribution, they produce state-of-the-art directions that are made available by the developers. \hl{Other than \mbox{\gz}, baseline methods are not available for the second user study. Instead we directly compared the disentanglement performance of \mbox{\gr}, \mbox{\gz}, and the combination of both.}

\hl{\textbf{Calibration}.} One of the challenges was to calibrate the strength of the directions across methods. The individual strengths of the directions needed to be adjusted per method and direction so the faces' changes were comparable. \hl{For the first user study,} we followed an approach similar to~\cite{khodadadeh2022latent}. We leveraged VGG-Face~\cite{parkhi2015deep} to find the smallest and largest limits of the applicable strengths where the faces could still be detected. We then divided this range into five intervals and used the resulting six images as the edited images. In total, we used 1000 reference images for the three aforementioned directions. Therefore, for each method, we had $1000$ reference images $\times3$ (coarse-grained) tasks per reference image $\times6$ resulting images per task $=18000$ 
total number of images for the analysis. We applied the same principle to the directions found by our participants. \hl{For the second user study, we applied the same principle where a simple \textit{dog detector} is used which is trained with Kaggle's Dog dataset~\cite{dogs-vs-cats}.}


{\bf Analysis}.
We re-trained InterFaceGAN for StyleGAN2 since it was originally released for StyleGAN. The directions of GANSpace\footnote{\href{https://github.com/harskish/ganspace}{https://github.com/harskish/ganspace}} and StyleFlow\footnote{\href{https://github.com/RameenAbdal/StyleFlow}{https://github.com/RameenAbdal/StyleFlow}} are already available for StyleGAN2 on Github. \hl{For the editing human faces user study,} we implemented \gz's scatter/gather functionality to find directions, \hl{since participants did not interact with \mbox{\gz} in the first user study.} Due to the nature of unsupervised direction discovery, not every direction can be found by the baselines. Coarse-grained tasks (glasses, smile and age) are all supported by the baselines and they are used in our analyses to compare \gr with the baselines for disentanglement performance. 
We also ran our analyses on fine-grained tasks and reported the metrics, even though they were not comparable with the baselines.
\hl{For the second user study, we trained a FastGAN using the AFHQ Dog dataset and extended the code of \mbox{\gz}. The details of the user study can be found in \S~\ref{section:cat-memes}.}

{\bf Facial Identity}. Although facial identity is used as a measure of disentanglement in the literature, it suffers from certain entanglement types. A subtle entanglement (\eg bigger eyes) does not change the face as much as a global entanglement (\eg getting older). As a result, the facial identity metric should not be compared across tasks.
We used a different face recognition model for facial identity analysis (FaceNet~\cite{schroff2015facenet}) than the model that is used for calibration (VGG-Face). Because the calibration step can bias the facial identity similarity metric if they use the same model. First, we extracted latent vectors from the last layer of FaceNet for the reference images. Next, we extracted latent vectors from the six edited images that originated by calibration. Then, we calculated the cosine similarity between the reference latent vectors and the six edited latent vectors. We averaged the results across tasks. Cosine similarity is between zero and one. Higher values represent a closer match between the vectors and therefore represent a more disentangled direction. 
The results can be seen in Table~\ref{tab:facial-identity}. As it can be seen \gr and InterFaceGAN have better facial identity retaining compared to GANSpace and StyleFlow. \gr outperforms all the baselines. The values also differ across tasks. For example, age has lower values compared to glasses and smile. This can be explained by images going through more major changes with the age direction making it harder to retain identity. For the fine-grained tasks: lipstick, eye, and curliness the \gr facial identity metrics are \hl{$.84 \pm .19$, $.85 \pm .21$, and $.73 \pm .22$} respectively. \hl{According to Mann-Whitney U test,} there are no statistically significant differences between coarse and fine-grained tasks. \hl{According to Friedman-Nemenyi test, only the age direction has statistically significant difference after Bonferroni correction (p=.03)}, which indicates age direction changes the face more than the other directions. 

\begin{table}[]
\begin{tabular}{lccc}
\toprule
 Coarse-Grained              & Glasses & Smile & Age  \\
\midrule
{{\sc interfaceGAN}\xspace}  &   $ .74 \pm .11$ & $.78 \pm .13$ & $.60 \pm .21$  \\
{{\sc GANspace}\xspace}  &   $.65 \pm .21$& $.76 \pm .15 $ &  $.42 \pm .34$  \\
{{\sc StyleFlow}\xspace}  &   $.55 \pm .24$ & $.69 \pm .15$ & $.48 \pm .40$  \\
\gz & $.58 \pm .22$ & $.71 \pm .22$ & $.45 \pm .39$ \\
\gr & \hl{$\mathbf{.84 \pm .19}$} & \hl{$\mathbf{.86 \pm .18}$} & \hl{$\mathbf{.67 \pm .23}$} \\
\bottomrule

\end{tabular}
\caption{Facial identity metrics of \gr with baselines IFG, GS, and SF for wearing glasses, smiling, and increasing age tasks. Higher values indicate higher disentanglement. 
}

\label{tab:facial-identity}
\end{table}

{\bf Classifier-Based}. Although facial identity is a useful metric for disentanglement, it does not entirely measure disentanglement. A face can go through minimal changes or the direction can be subtle in which case the face retains most of its identity. Another way to quantify disentanglement is to use facial attribute classifiers and measure the cosine similarity of their latent vectors after they are edited. First, we can extract the latent vectors of the edited and the reference images with a classifier that is trained for the goal of the direction. For example, if the direction is age, we can extract the latent vectors of the reference images and the edited images using an age classifier. Then, we can calculate the cosine distance between the vectors coming from the age classifier. We can do the same calculation using different face attribute classifiers such as baldness, hair color, face roundness, facial hair, etc. Since a disentangled direction should not change other facial attributes, it is expected to have a higher cosine similarity when different face attribute classifiers are used compared to the age classifier. However, in practice, it is costly to train many different facial attribute classifiers. Instead, we used an open-source face attribute classifier model called FAN~\cite{he2018harnessing} that can detect 40 binary attributes with one model including all six tasks in our study. FAN takes an image as an input and outputs a vector of size 40 that consists of 1s and 0s indicating whether that attribute is present or not in the image. We used FAN and recorded: \one whether the targeted attribute (old) was detected, \two percentage of attributes that were lost (out of 40), and \three percentage of new attributes (out of 40) that were detected after the direction is applied. We averaged the results across all the coarse-grained tasks and the results can be seen in Table~\ref{tab:classifier-based}. Ideally, after the image is edited, the output of FAN should not lose any attribute, and it should not find any new attributes other than the targeted attribute. 

\begin{table}[]
\begin{tabular}{lccc}
\toprule
  Coarse-Grained             & Success (\%) ($\uparrow$) & Lost (\%) ($\downarrow$) & Found (\%) ($\downarrow$)  \\
\midrule
{{\sc interfaceGAN}\xspace}  &   $72.34 \pm 18.15$ & $6.32\pm 3.64$  & $8.82 \pm 4.12$  \\
{{\sc GANspace}\xspace} &   $69.68 \pm 21.64$ & $8.74 \pm 6.61$ &  $10.12 \pm 7.12$ \\
{{\sc StyleFlow}\xspace} &   $69.13 \pm 27.16$ & $11.98 \pm 6.85$ & $13.87 \pm 6.91$ \\
\gz & $66.99 \pm 35.98$ & $12.12 \pm 7.35$ & $10.33 \pm 5.22$ \\
\gr & \hl{$\mathbf{74.81 \pm 12.66}$} & \hl{$\mathbf{3.41 \pm 4.33}$} & \hl{$\mathbf{4.59 \pm 4.12}$}\\
\bottomrule

\end{tabular}
\caption{Facial attribute classifier metrics of \gr with baselines IFG, GS, and SF for wearing glasses, smiling, and increasing age tasks. Success percentage indicates how successful the direction is in adding the target attribute when applied. The lost percentage indicates how many facial attributes are lost when the direction is applied. The found percentage indicates how many facial attributes are introduced when the direction is applied. Lower values for lost and found indicate higher disentanglement.
}
\label{tab:classifier-based}
\end{table}

\hl{According to Friedman's test, after Bonferroni correction, there is no statistically significant difference in success rates. This indicates that} all methods introduce the targeted attribute with similar percentages. However, InterFaceGAN and \gr have statistically significant lower lost and found attributes which is a result of their disentanglement capabilities (p=0.04 and 0.02 respectively). \gr outperforms all the baselines.  For the fine-grained tasks: the \gr success, lost and found metrics are \hl{$63.91 \pm 23.18$, $6.00 \pm 7.91$, and $5.61 \pm 11.23$} respectively. \hl{According to Mann-Whitney U test,} there are no statistically significant differences between coarse and fine-grained tasks. \hl{According to Friedman-Nemenyi test, only the success metric has statistically significant difference (p=.02) after Bonferroni correction. This can be explained by fine-grained tasks being more subtle} by definition. That being said, lost and found metrics are similar which again shows that the directions are disentangled regardless of the task.

\hl{\textbf{Dog Breed}}. \hl{In order to measure disentanglement in the second user study, we used a dog breed classifier which is trained with the Oxford Dog dataset~\cite{parkhi2012cats}. Similar to the facial identity metric in the first user study, a disentangled direction is expected to preserve the breed more. It should be noted that human faces are better studied and have better models (like FAN) than dogs. As a result, the analyses in this section should be viewed as complementary to the previous disentanglement performance analysis.} 

\hl{
In the second user study, we asked participants to use \mbox{\gr}, \mbox{\gz}, and the combination of both (\mbox{\gz}+\mbox{\gr}) to create dog memes. In the last task, participants disentangled directions that they already found with \mbox{\gz}. Similar to facial identity metric, we extracted latent vectors of the reference images and the edited images and calculated cosine similarity for each task. Cosine similarities are between zero and one, where higher values represent a more disentangled direction. The results for \mbox{\gr}, \mbox{\gz}, and the combination of both (\mbox{\gz}+\mbox{\gr}) are $.91 \pm .12$, $.45 \pm .32$, and $.90 \pm .15$, respectively. According to Friedman-Nemenyi test, across three tasks, only \mbox{\gz} has statistically significant difference (p=.02) after Bonferroni correction. This shows that \mbox{\gz} is significantly worse at creating disentangled directions than \mbox{\gr}. More interestingly, participants were able to disentangle the directions they discovered with \mbox{\gz} using \mbox{\gr}. This analysis highlights how \mbox{\gr} can be either used by itself or complementary to other direction discovery methods for disentangled direction discovery.
}

\subsection{Iterative disentanglement}
\label{section:disentanglement-patterns}

User-driven disentanglement is one of the biggest contributions of \gr. Previously in \S~\ref{section:disentanglement-performance}, we showed the disentanglement performance of the directions that users found at the end of each trial. In this section, we analyze how the disentanglement changes over time as the participants interact with \gr. We provide insight into interactive disentanglement in using classifier-based disentanglement over time.

For each trial when the users selected exemplary images and tested a direction for the first time, we saved the direction as the `entangled' direction. After they interacted with \gr to improve the direction, we saved the final version as the `disentangled' direction. Previously in \S~\ref{section:disentanglement-performance}, we reported the results for the `disentangled' direction. We did the same analysis for the `entangled' direction. Then, we subtracted the success, lost, and found percentages of the `disentangled' direction from the `entangled' direction and reported it for each trial. Positive values in success imply the \textit{final} direction introduces the target facial attribute more often than the \textit{initial} direction. Whereas, negative values in lost and found imply that the \textit{final} direction is more disentangled than the \textit{initial} direction. The results can be seen in Table~\ref{tab:classifier-based-difference}. 

\begin{table}[]
\begin{tabular}{lccc}
\toprule
Over Time               & Success (\%) ($\uparrow$) & Lost (\%) ($\downarrow$) & Found (\%) ($\downarrow$)  \\
\midrule
Glasses  &   \hl{$ 0.29 \pm 1.38$} & \hl{$ -7.61\pm 4.51$}  & \hl{$ -5.80\pm 4.73$}  \\
Smile &   \hl{$ 0.60 \pm 0.69$} & \hl{$ -4.69\pm 6.54$}  & \hl{$ -3.02\pm 3.51$}  \\
Age &   \hl{$ 0.38 \pm 1.01$} & \hl{$ -6.95\pm 7.82$}  & \hl{$ -4.00\pm 5.51$}  \\
Lipstick & \hl{$ 1.08 \pm 1.11$} & \hl{$ -7.95\pm 7.99$}  & \hl{$ -6.12\pm 7.65$}  \\
Eye & \hl{$ 1.37 \pm 0.92$} & \hl{$ -6.57\pm 4.66$} & \hl{$ -7.69\pm 8.10$}\\
Curliness & \hl{$ 0.58 \pm 1.56$} & \hl{$ -5.89\pm 5.99$}  & \hl{$-5.78 \pm 6.11$}  \\
\bottomrule

\end{tabular}
\caption{Facial attribute classifier metrics of \gr for each trial when the final `disentangled' direction results are subtracted from the initial `entangled' direction.  Positive values in success indicate improvement in the target facial attribute over time. Negative values for lost and found indicate higher disentanglement over time.}
\label{tab:classifier-based-difference}
\end{table}

\hl{According to Mann-Whitney U test,} there is a statistically significant difference between the initial and final directions for both loss and found percentages \hl{(p = 0.02, 0.04, 0.02, 0.02, 0.02, and 0.03 respectively for lost) (p = 0.03, 0.03, 0.04, 0.02, 0.03, and 0.03 respectively for found).} As the users interacted with \gr, they disentangled the initial direction consistently for each trial. However \hl{according to the Mann-Whitney U test,} there is no statistically significant difference between the initial and final directions for the success percentages. In other words, the target facial attribute is present when the initial direction is applied as well as the final direction. \hl{The improvement is in the disentanglement performance over time.} 

\hl{We also analyzed how the disentanglement performance improved over time. Every time the participants applied a disentanglement, global or local, we extracted the current direction and ran the same analysis we ran earlier in this section. The results can be seen in Figure~\ref{fg:disentanglement_over_time}. As it can be seen, the disentanglement performance increases over time (percentages get lower). Interestingly, on average it took $4.12$ and $5.32$ actions for participants to reach at least $90\%$ of the maximum disentanglement they could achieve. This can be explained by participants focusing on global disentanglement at the beginning which is easier to observe with the metrics, since global disentanglement change the image more than local disentanglement. $73.9\%$ of the first $5$ actions consist of global disentanglement, which supports the previous observation.
}

\fg{disentanglement_over_time}{disentanglement_over_time}{0.9}{Lost and found percentages over user actions when the current direction metrics are subtracted from the initial entangled direction metrics. Lower values indicate higher disentanglement. The shaded regions represent 0.2 times the standard deviation to make plots readable. The values of tasks should not be compared with each other as discussed.}

\subsection{User behavior}
\label{section:user-behavior}

In this section, we report how the users interact with \gr. Overall, the average time to complete a coarse and fine-grained task were \hl{8 minutes 54 seconds and 8 minutes 29 seconds} respectively. \hl{For the second user study, the average time to create a dog meme was 9 minutes and 12 seconds for \mbox{\gz}, 9 minutes and 21 seconds for \mbox{\gr}, and 6 minutes and 42 seconds for disentangling the directions of \mbox{\gz} using \mbox{\gr}. The faster times in the last task can be explained by participants not needing to choose exemplary images. The results show that using \mbox{\gr} from scratch to find a disentangled direction takes less time than disentangling an entangled direction that is found with \mbox{\gz}, including the time that it takes to find a direction with \mbox{\gz}. However, it can also be argued that algorithm-driven approaches can find a direction significantly faster than a user-driven tool like \mbox{\gz}. Therefore using algorithms in combination with \mbox{\gr} for disentanglement can yield better results. We did not analyze the optimal disentanglement procedure in this work and left it as future work.} Participants spent \hl{$32.32\%$}  of their time selecting exemplary images, \hl{$46.34\%$} on live-testing the directions, \hl{$7.66\%$}  on highlighting, \hl{$8.17\%$} on weight adjustments, and the remaining \hl{$5.51\%$} on applying masks. The average number of positive or negative examples selected per trial is \hl{$15.32$}. The average number of highlighting per trial is \hl{$2.75$}. The average number of times the weights of the exemplary images are adjusted per trial is \hl{$8.54$}. The average number of directions tested on live-testing per trial is \hl{$3.72$}. However, participants tested each direction thoroughly and changed the individual weights of the test images on live-testing \hl{$8.56$} times per direction. The average number of masks the participants tested per trial is \hl{$2.53$}. The average number of times the participants applied the mask to all the test images is \hl{$1.82$} per trial.

Participants were able to find state-of-the-art disentangled directions from scratch in under $10$ minutes \hl{$81.19\%$} of the time. This shows how \gr can \hl{efficiently enable the end user to find disentangled directions} when a dataset or a classifier is not available. Moreover, participants got better at disentanglement as they spent more time with \gr. \hl{For the first user study,} the initial directions they found were \hl{$41.33\%$} more disentangled (according to classifier-based metrics) after the first three trials which is statistically significant (p=0.04) \hl{according to to Mann-Whitney U test}. They achieved this by spending \hl{$13.1\%$} more time on the initial image selection which is also statistically significant (p=0.04) \hl{according to to Mann-Whitney U test}.

\section{Qualitative Results}


\fg{user_responses}{participant_rating}{1}{The participants' average ratings. The questions are explained in \S~\ref{subsec:measurement}. All questions used a seven-point Likert scale.}

We employed a method akin to the Affinity Diagram approach~\cite{holtzblatt1997contextual}, and we aggregated participants’ responses. We summarized their perceived ease, the perceived success of disentanglement, and the perceived success of finding the directions using \gr in \S~\ref{subsec:overall}. We also report the cognitive load responses using the
mental demand, effort and frustration dimensions in the NASA TLX questionnaire in \S~\ref{subsec:tlx}. Additionally, we extracted recurring themes regarding how participants assess the usefulness of the individual components of \gr in \S~\ref{subsec:ablative}. Specifically, the first author transcribed participants’ responses to develop the initial codes, which were then reviewed by the second author. Disagreements were resolved via discussion between the two authors. Figure~\ref{fg:user_responses} shows the \hl{average} ratings of the participants for \gr on ease of use, perceived disentanglement success, perceived trial success, cognitive load, and ablative assessment of each component's usefulness. 

We also show some qualitative images showing the disentanglement performance of \gr \hl{for the first user study}. In \fgref{gr_vs_others}, we compare the resulting images of \gr with the baselines. As can be seen, \gr has better disentanglement. Additionally, we show directions that participants found during various trials in \fgref{participant_qual} (which uses the same reference image as \fgref{gr_vs_others}). As can be seen, the participants have successfully disentangled directions in all of the tasks. In \fgref{disentanglement}, we show the improvement of the direction `glasses' as the participant interacts with \gr. As can be seen, the participant disentangles age and gender with global disentanglement. Then, the participant disentangles the remaining local attributes with local disentanglement. \hl{For the second user study, we show some dog memes the participants created in Figure~\ref{fg:dog_memes}. In Figure~\ref{fg:dog_user_study_comparison}, we also show some directions the participants found using \mbox{\gr}, \mbox{\gz} and both. As it can be seen, the entanglement issues improved after the participants used \mbox{\gr}.}


\fg{gr_vs_others}{updated_coarse}{1}{Comparison of state-of-the-art direction discovery methods and \gr. The reference row is the original image. \gr has better disentanglement than other methods.}

\fg{participant_qual}{participant_qual}{1}{Participants found various disentangled directions using \gr. Each image is generated from a disentangled direction, found by the participants. Each column is for a different trial. Participants successfully disentangled the directions.}

\fgw{disentanglement}{disentanglement}{1}{Improvement of the direction glasses as the participant interacts with \gr. The `entangled' direction has entanglement with age, gender, and other various local attributes. After the global disentanglement, the direction is disentangled from age and gender but still entangled with local attributes (\eg hairstyle, mouth, etc.). Finally, after the local disentanglement, the glasses direction is disentangled.}

\fgw{dog_memes}{dog_memes}{0.6}{Dog memes created by the participants. A disentangled edited image makes a higher quality meme.}
\fgw{dog_user_study_comparison}{dog_user_study_comparison}{0.6}{Comparison of different tasks in the dog meme user study. \gz creates entangled directions which result in worse quality edited images. \gr can disentangle a direction that is found with \gz.}

\subsection{Overall assessment}
\hl{P1-16 represents the 16 participants from the first user study. P17-32 represents the 16 participants from the second user study.}

\label{subsec:overall}
\subsubsection{Ease of using the tool}
All of the participants except (P9, \hl{P17, and P26}) gave a rating equal or greater than five, when they were asked to rate how easy \gr was to use. For example P1 commented on the intuitive workflow of \gr: ``The steps are really intuitive. You just select images that have the target attribute and then highlight the region if the results are entangled''.
P3, P7, \hl{P17, P11, P14, P30, and P31} commented on how fast and easily they could find the directions.
P4, P5 \hl{and P22} pointed out that they had to fine-tune the individual strengths on some of test images for the target attribute to appear, which made it harder to access the directions easily. \hl{When the directions are entangled, they can fail to work on some test images. For example P22 said: ``Initially, the direction was not working on all of the dogs. When I disentangled the direction, it started to work''.}  
P1, P2, \hl{P23, P12, and P32}  mentioned it was intuitive to figure out which images to select for global entanglement.
P9 gave below-five rating (four) and mentioned the lack of guidance in the tool: ``When I got a direction that I did not expect, the next step to take was not always clear''. \hl{P17 and P26 shared similar concerns.
Ratings of P9, P17 and P26 (below-five)} are outliers based on the IQR analysis. \hl{Overall, participants thought the tool was easy to use.}

\subsubsection{Perceived success of disentanglement}
All of the participants except P8 \hl{and P29} gave a rating greater than four, when they were asked to rate their perceived success of disentanglement. 
P9, who previously rated four for the ease of use, thought the disentanglement was successful: ``I could get rid of most of the entanglement problems as I changed the weights''. \hl{P26, who also rated four for the ease of use, said: ``I could clearly see the disentanglement when I improved the image that I found before with \mbox{\gz}.''}
P8, who rated three, thought that it was hard to preemptively avoid entanglement and the resulting directions were not always as they were envisioned.  P8's and \hl{P29's} ratings are outliers based on the IQR analysis.
P1, P10, \hl{P15, P23, and P31} pointed out that they could disentangle the directions better as they got more experience with \gr.
Participants also pointed out variations in entanglements: ``Some entanglements were more obvious, for example, gender and age. I tried to disentangle them first'' (P3) and different strategies to overcome them: ``Some entanglements were more subtle such as gaze direction or chin roundness. I tried to ignore those and focus on the more obvious ones first'' (P5).
\hl{``Sometimes I could see the breed of the dog changing. After balancing out the exemplary dogs, It got much better.'' (P21). Overall, the participants thought they were successful in disentangling the directions which is also backed by the quantitative analysis.} 

\subsubsection{Perceived success of the trials}
All of the participants gave a rating of six or seven, when they were asked to rate their perceived success in finding the direction for trials.
Participants had different reasons why they were confident with their directions: ``I could see that the direction was working on all of the test images'' (P1), ``I could build a diverse set of exemplary images in all trials, so the resulting directions were successful'' (P3), and ``I could see the improvement in directions after I spend some time on them, progression was convincing'' (P10).
Some participants felt successful after changing the strength of the direction and saw the transition from the initial image (P2, P7, P8, \hl{P12, P16, P20, P30). P17 and P32 pointed out the open-ended nature of creating memes and commented that they were surprised that they could find the directions they envisioned while creating the memes. P29 mentioned how he adjusted for the tool: ``It was not possible to find everything I was looking for, instead I got inspired by what I could find.''. Overall, participants felt like they were able to finish the tasks.}

\subsection{Cognitive load by NASA TLX}
\label{subsec:tlx}
Only one participant (P9) gave a rating higher than four (neutral) for the mental demand dimension and pointed out that the main mental demand was to come up with ways to disentangle a direction when the entanglement was not very clear. As explained by P9: ``Sometimes I could see there were some entanglement problems with the direction, the face looked like someone else, but I did not know how to fix it''. 
P9 also gave a rating higher than neutral for the frustration dimension citing the same reasoning. Rating of P9 on frustration is an outlier based on the IQR analysis. Most of the participants found the tasks were not mentally demanding or frustrating as mentioned by \hl{P18:} ``It was easy and a lot of fun''. P8: ``It behaved as I expected and that was really satisfying''. P4, P7 \hl{P13, and P31} mentioned that they could see their progress over time and were motivated by it.
P8 was the only participant who rated higher than neutral for the effort dimension and said: ``For some trials, I had to do multiple iterations until I was confident''. 
Rating of P8 on the effort is an outlier based on the IQR analysis. 
The rating of P4 for the frustration dimension was five because of how the directions could behave unexpectedly. As mentioned by P4: ``Sometimes the directions did not work on all the test images or the directions required specific strengths to appear. I had to test a lot of different values in some trials''. Rating of P9 (five) and P4 are outliers based on the IQR analysis. Most of the participants were not frustrated, in fact, they ``enjoyed'' (P3, P7, \hl{P17, P18, P19, P21}) the tasks.

\subsection{Ablative Assessment}
\label{subsec:ablative}

In this section, we summarized the recurring themes based on participants' responses on the usefulness of individual components of \gr.

\subsubsection{Participants could preemptively avoid entanglement issues by curating a set of positive and negative examples.} 
Some participants (P1, P2, P5, P7, P10, \hl{P13, P17, P19, P21, P22, P24, P25, P28-32}) pointed out that they got better at disentanglement as they learned more about the model and more about the entanglements in the directions. As P1 mentioned: ``After the first couple of trials, I was looking to avoid some entanglements proactively. I was trying to balance out concepts such as gender, age, and glasses by choosing a diverse set of examples''. Participants also had different strategies to avoid entanglement, for example as P2 mentioned: ``For the curliness direction, I had a hard time finding male examples. So instead, I selected most of my positive and negative examples as females so the final direction was not entangled with gender''. \hl{Another example was from P23: ``I was selecting just a couple of exemplary images to see the initial entanglement and then try to select the next example based on the entanglement. From there, I added examples one-by-one''}. P5 pointed out that creating a `balanced' set requires tuning through trial and error: ``It was not as easy as selecting the same number of female and male positive examples to disentangle gender. It required tuning''. 
P7, P10, \hl{P23, P27, and P31} mentioned that with more available time, they could tune out the entangled features more.

\subsubsection{Participants struggled with entangled directions that do not introduce the target facial attribute on all the test images.} 

P4, P5, \hl{P11, P13, P15, P23, and P29} pointed out the lack of consistency in the live-testing area when the direction is entangled. Specifically, they talked about the cases when a direction did not work on certain test images, but worked on others, as mentioned by P5: ``When a direction did not work on a couple of test images, it was hard to believe in the direction''. 
P4 talked about the same issue with more nuance: ``Because of the entanglement, lipstick direction did not work on all the test images. But as I disentangle the direction and make the direction purer, it started to work on more images''.
\hl{P15 also mentioned the same issue: ``Some directions changed the breed of the dog but not for all of the test images. It was a little confusing''.}
To our surprise, some participants were more positively reinforced when the direction worked well on a couple of test images but not on all of them. They thought that it was ``not possible'' (P1 and \hl{P19}), ``a limitation of the model'' (P2 and \hl{P32}), and they tried to ``get right'' as many images as they can (P9, \hl{P13, and P22}). 
Similarly, some participants mentioned the lack of quantitative metrics in the live-testing area (P3, P10, \hl{and P25}), as mentioned by P10: ``I wish there were metrics at testing, so I knew I was improving the direction more objectively''. 
P6 \hl{(and P31)} had a different strategy: ``I focused on the worst looking test examples and tried to improve them. The rest of the test examples usually followed''.

\subsubsection{Some participants favored one of the disentanglement approaches over the other.} 

To our surprise, some participants found the global disentanglement more intuitive, as mentioned by \hl{P21: ``Changing the weights to improve images were really helpful, I did not use the masking a lot.''} and P5: ``Changing the weights of selected images was helpful whenever disentanglement was necessary and the results were as expected''. 
Similarly, P1 preferred to use changing weights over masking, saying: ``I exactly knew what `changing weights' did, but I was not sure how masking worked under the hood''. However, some participants preferred masking, as mentioned by P2: ``Masking is easy to use, I just highlighted the area of interest and the directions were disentangled''. P7 \hl{and P29} mentioned masking was ``faster'' and ``easier'' than changing weights to disentangle the direction. Some participants utilized both and even created their own system, as mentioned by P3: ``If the entangled features were obvious such as gender or age, I fixed them with changing weights. For more subtle entanglements, I used masking''. All of the participants gave ratings of above four to the \textit{changing weights} component.

\subsubsection{Participant used local disentanglement with various masks.} 

Masking was used in various ways by some participants, as P2 mentioned: ``Whenever the lighting on the face was changing, I highlighted the entire face and used the masking to reduce lighting entanglement'' (using the discard feature). Some participants highlighted more subtle changes such as cheeks (P3) or forehead (P4). However, according to P4, subtle masks did not always work and bigger masks were more ``consistent''. P5 gave a more nuanced explanation of why masking was harder to use in certain situations: ``Trying to disentangle age from gender was harder with masks because both directions were not contained to an area''. Some participants did not use the \textit{discard feature} of the masks to eliminate entanglement (P8, P9, and P10) but instead used the \textit{preserve feature} which is explained in \S~\ref{subsec:local}. For example, P8 said: ``For the eyeglass trial, I chose the best test example that had glasses. Then I used the masking on the glasses and kept them (preserve). The direction got significantly more disentangled''. P6 only used the \textit{discard feature} to improve the ``worst-looking'' test examples. \hl{A similar trend can be seen in the second user study, where participants had different strategies. For example P26 said: ``Masking the ears were resulting in change of the breed, so instead I focused on the discard feature''. P17 and P21 mentioned `using different combination of masks' to end up with a disentangled direction.}

\section{Discussions}

This section discusses several issues in the current tool and possible solutions for future work. 

{\bf Limitations of the current study}. First of all, future work can increase the number of participants beyond the current user studies. Moreover, we can also use a narrower user group (\eg artists trying to use GANs for creative purposes). In the `age' trial of the first user study, participants had various interpretations of the task, \eg some participants found directions for wrinkly faces, some found white hair, etc. This trial can be improved so that it is easier to compare it with prior work.

{\bf Improving disentanglement with prior analysis}. 
Currently, \gr is designed to iteratively disentangle a direction. Typically, the user identifies an entanglement problem and then tackles it with either global or local disentanglement. While fixing one type of entanglement, the disentanglement process can introduce a new type of entanglement. For future work, one possible idea is to analyze entanglements in the model via an unsupervised algorithm prior, such as clustering. This would allow the model to predict which type of entanglements appear together and can be used to feed-forward information about the entanglements while the users interact with the system. 

{\bf Providing guidance through exemplary image selection}.
Image selection plays an important part in global disentanglement. Currently, the user goes through the image gallery which consists of randomly sampled images. We decided to go with random sampling because, in an earlier version of \gr, user recommendations for selections created bias and entanglement in the resulting direction. In the future, non-biased guidance through image selection can be implemented to help the users. This can be achieved by a text prompt indicating what the user is searching for (\eg young blond males) or if the search is about more subtle attributes, it can be through highlighting a region.

{\bf Providing guidance and metrics on disentanglement}.
In the future, more guidance on disentanglement can be provided to the user. This can be achieved with heat maps that show the changed regions as well as metrics (\eg facial identity) that indicate the disentanglement performance. As another solution, the user can indicate when a disentanglement works and does not work, which can then be used to learn the correlations in the GAN model. The information can then be used for further guidance as the users disentangle a direction. As a result, the tool can adapt to user feedback.



\section{Conclusion \& Future Work}
\hl{In this section, we summarize key insights of \mbox{\gr} to help future work.}

\begin{itemize}

\item \hl{Curating a balanced set of images to find a disentangled direction is a trivial task in disentanglement. Surprisingly, as the participants interacted with \mbox{\gr}, they got better at avoiding the initial entanglement issues.}

\item \hl{Directions can fail to generalize across test images due to entanglement. But this failure can also be caused by the limitation of the model. In other words, some test images could not be successfully edited with the direction. This phenomenon was confusing to the users since they did not know if their direction was entangled or it was a limitation of the generative model. In the future, figuring out the model limitations automatically and communicating it to the user can become an important task.}

\item \hl{A future research direction can be figuring out how to utilize algorithmic direction discovery methods in user-driven direction discovery. In the second user study, we discovered that starting with a direction can speed up disentanglement. Using algorithm-driven direction discovery methods to initialize the search process or leveraging them throughout the human-AI interaction is left as future work.}

\item \hl{As the participants disentangled directions, they learned the common entanglement issues specific to the generative model. With each new task, they had to disentangle the same issues such as age and gender. In the future, an algorithmic way to get rid of this repetitive task can be investigated.}

\item \hl{Subtle entanglement issues are harder to measure as discussed in \S~\ref{section:disentanglement-patterns}. In the future, new metrics can be investigated specifically tailored for local entanglement issues.}

\item \hl{Throughout the workflow of \mbox{\gr}, selecting exemplary images was time consuming as discussed in \S~\ref{section:user-behavior}. Users were `communicating' with the model through selecting images. For example by selecting images with blonde hair, the users were interacting with the model to find the respective direction that changes the hair color. In prior work, instead of exemplary images, StyleCLIP~\cite{Patashnik_2021_ICCV} leverages text prompts for the same interaction, which is faster than selecting images. However the downside of natural language is, it does not have the fine-grained optimization a set of images can provide. In the future new interaction methods should be investigated that provides both fast and detailed communication between the users and AI models.}

\end{itemize}


\bibliographystyle{ACM-Reference-Format}
\bibliography{ref_main}

\end{document}